\documentclass[prd,aps,showpacs,nofootinbib]{revtex4}
\usepackage{amssymb}
\usepackage{graphicx}
\usepackage{amsmath}

\newcommand{\be}{\begin{equation}}
\newcommand{\ee}{\end{equation}}
\newcommand{\bq}{\begin{eqnarray}}
\newcommand{\eq}{\end{eqnarray}}

\begin{document}

\title{Invariance of the fine structure constant with temperature of the expanding universe} 
\author{Cl\'audio Nassif* and A. C. Amaro de Faria Jr.**\\
e-mail: {\bf cnassif@cbpf.br*},~{\bf antoniocarlos@ieav.cta.br*}}
\altaffiliation{{\bf CBPF}: Centro Brasileiro de Pesquisas F\'isicas, Rua Dr. Xavier Sigaud 150, Urca, CEP: 22.290-180, Rio de
Janeiro-RJ, Brazil.\\
{\bf **UTFPR-GP}: Federal Technological University of Paran\'a, Avenida Professora Laura Pacheco Bastos, 800, Bairro Industrial, 
CEP: 85.053-510, Guarapuava-PR, Brazil.\\
{\bf **IEAv}: Institute of Advanced Studies, Rodovia dos Tamoios, Km 099, CEP: 12.220-000, S\~ao Jos\'e dos Campos-SP, Brazil.}

\begin{abstract}
Our goal is to interpret the energy equation from Doubly Special Relativity (DSR) of Magueijo-Smolin with an invariant Planck energy
scale in order to obtain the speed of light with an explicit dependence on the background temperature of the expanding universe\cite{1}.
We also investigate how other universal constants, including the fine structure constant, have varied since the early universe and, thus, 
how they have evoluted over the cosmological time related to the temperature of the expanding universe. For instance, we show that both the
Planck constant and the electron charge were also too large in the early universe. However, we finally conclude that the fine structure 
constant has remained invariant with the age and temperature of the universe, which is in agreement with laboratory tests and some 
observational data. 
\end{abstract}


\maketitle

\section{Introduction}

 There are many theoretical proposals for variation of the fundamental constants of nature, including the variation of the fine structure
 constant $\alpha$\cite{2}\cite{3}\cite{4}\cite{5}\cite{6}. Furthermore, many evidences behind recent claims of spatial variation in the
 fine structure constant, due to ground-based telescopes for the observations of ion absorption lines in the light of distant
 quasars, have led to much discussion because of the controversial 
 results about how different telescopes should observe distinct spatial variations on $\alpha$\cite{7}. Variation over the cosmological time
 has also been conjectured\cite{8}\cite{9}\cite{10}. In view of all this, we should be careful to investigate the veracity of such
 controversial results. For this, we will start from a new interpretation of the well-known Magueijo-Smolin (MS) dispersion relation by
 taking into account the presence of an isotropic thermal background field with temperature $T$, which has been addressed in a previous 
 Brief Report\cite{1}, where we have found the dependence of the speed of light with temperature of the expanding universe. Starting from
 this result\cite{1}, the present work goes further in order to obtain the variation of the Planck constant with temperature and also the
 variation of the electron charge with temperature of the universe; however we finally conclude that the fine structure constant $\alpha$,
 as a dimensionless number, has remained invariant with the cosmic time scale (temperature). Thus, we will show the invariance of $\alpha$,
 i.e., we will find $\alpha^{\prime}=\alpha(T)=\alpha=q_e^2/4\pi\epsilon_0\hbar c$, where $\alpha^{-1}\approx 137.035999037(91)$\cite{11}. 

  It is important to mention that our result about the invariance of $\alpha$ is in agreement with the observational results of J. Bahcall, W. Sargent and 
 M. Schmidt\cite{12} who measured the fine structure constant in quasar $3$C$191$ and showed that its value did not vary significantly
with time, giving support to our theoretical result. Furthermore, we should mention that a recent article on the Bayesian reanalysis of the 
quasar dataset\cite{13} reveals significant support for a skeptical interpretation in which the apparent dipole effect is driven solely 
by systematic errors of opposing sign inherent in measurements from the Keck and VLT telescopes employed to obtain the observations\cite{7}. 
Thus, this reanalysis leads us to question such results which show that the fine structure constant exhibits spatial variations. This 
strengthens our defense in favor of its isotropy and also its invariance, according to Occam's razor\cite{13}. 

 Although recent astrophysical data suggest that the fine structure constant $\alpha$ has increased over the cosmological time, where the
combined analysis over more than 100 quasar systems has produced a value of a relative change of $\Delta\alpha/\alpha=-0.57\pm 0.10
\times 10^{-5}$, which is at the $5\sigma$ significance level\cite{14}, in contrast, we have laboratory tests that cover only a short 
time span and they have found no indications for the time-variation of $\alpha$\cite{15}. Their advantage, however, is their great
accuracy, reproducibility and unequivocal interpretation.

 \section{A new reading of MS-DSR energy equation in cosmological scenario}

Magueijo and Smolin(MS)\cite{16} proposed a DSR theory, where the total energy of a particle should be corrected close to the Planck energy
scale ($E_P$) as being an invariant scale, so that they have obtained

\begin{equation}
E=\frac{mc^2}{1+\frac{mc^2}{E_P}}, 
\end{equation}

where $mc^2=\gamma m_0 c^2$, being $\gamma=1/\sqrt{1-v^2/c^2}$. $E_P$ $(=M_P c^2 )$ is the Planck energy scale ($\sim 10^{19}$ GeV) 
and $M_P$ $(\sim 10^{−4}$g) is the Planck mass.

Notice that, if $E_P > 0$, the energy of a particle is smaller than the usual $E = mc^2$ ; however if $E_P<0$,thus MS-energy\cite{16}
becomes larger than the usual $mc^2$ and, in fact, diverges for Planck mass particles ($E_P$). For reasons that will be naturally 
justified later, let us consider the 2nd case ($E_P<0$)\cite{16}, where the energy of the particle diverges very close to the Planck
energy scale. So we have

\begin{equation}
E=\frac{mc^2}{1-\frac{mc^2}{E_P}}, 
\end{equation}
with $mc^2=\gamma m_0 c^2$.

The Eq.(2) above shows that the energy of the particle diverges for the Planck energy scale ($E_P$). Such a correction on energy close to
the Planck scale could be interpreted as being an effect of light dispersion that provides a variation of the speed of light with the
energy scale, where $c$ should increase for higher energy scales, so that we can write Eq.(2) as follows:

\begin{equation}
E=mc^{\prime 2}=mc^2(E), 
\end{equation}
where the speed of light depends on the energy scale, such that we have $c^{\prime}=c(E)\geq c$ according to Eq.(2). So, combining Eq.(2)
with Eq.(3), it is easy to see the following dispersion relation for light, namely:  

\begin{equation}
c^{\prime}=c(E)=\frac{c}{\sqrt{1-\frac{E}{E_P}}}, 
\end{equation}
from where, within a cosmological scenario, we can think that the speed of light has diverged in the early universe close to the Planck
temperature $T_P$, which is directly related to the Planck energy scale $E_P$ connected to a minimum length (Planck length $l_P$) as
being an initial singularity with Planck temperature, namely: 

\begin{equation}
E_p (\propto l_P^{-1})=M_Pc^2=K_B T_P,
\end{equation}
where $l_P(\sim 10^{-35}$m) is the Planck length, $M_P$ is the Planck mass and $T_P(\sim 10^{32}$K) is the Planck temperature in the 
early universe with radius of about $l_P\sim 10^{-35}$m. 

Combining Eq.(5) with Eq.(4) and considering any energy scale as being of the form $E=K_BT$ to represent the thermal background energy,
we find 

\begin{equation}
c^{\prime}=c(T)=\frac{c}{\sqrt{1-\frac{T}{T_P}}},
\end{equation}
where the speed of light has been varied with the temperature of the expanding universe, so that it has diverged in the early universe
when $T=T_P$ (Planck length $l_P$). 

We could think that the background thermal energy in the universe works like a special ``medium'' with a kind of anomalous index of
refraction [$n(T)$] in the sense that it must obey the following inequality $n\leq 1$, leading to an increase of the speed of light with
higher temperatures, i.e., in the early universe, we had $n=n(T)<<1$, as $T\approx T_P$ and, thus, $c$ has increased drastically at 
that time. In view of this reasoning, we can simply write Eq.(6) in the form $c(T)=c/n(T)$, where $n=n(T)(\leq 1)=\sqrt{1-\frac{T}{T_P}}$,
$n$ being zero for $T=T_P$, which leads to $c^{\prime}=c(T_P)=\infty$\cite{1}. Here we should stress that such a divergence is possible only if we had to
consider the 2nd case for the MS-DSR energy equation\cite{16}, which leads to a realistic result within a cosmological scenario, i.e., 
there must be a singularity for the Planck scale. 

Now, basing on Eq.(6), we can simply rewrite MS-energy equation [Eq.(2)] in cosmological scenario due to the background temperature, 
namely: 

\begin{equation}
  E(T)=\Gamma(T)mc^2=mc^2(T)=mc^{\prime 2}=\frac{mc^2}{\left(1-\frac{T}{T_P}\right)}, 
  \end{equation}
where we have $\Gamma(T)=1/n^{2}(T)=1/\left(1-\frac{T}{T_P}\right)$. The factor $\Gamma(T)$ has a non-local origin since it is related to
the background temperature of the whole universe. 

 From Eq.(7), we find $c^{\prime}=c(T)=\sqrt{\Gamma(T)}c=\gamma_{T} c$, with $\gamma_{T}=1/n(T)=1/\sqrt{1-T/T_P}$. So, the change in the 
 speed of light is $\delta c=c^{\prime}-c$, i.e., $\delta c=(\gamma_{T}-1)c=(1/\sqrt{1-T/T_P}-1)c$. For $T<<T_P$, we get
 $\delta c\approx 0$.

We should stress that Eq.(7) provides a correction on energy $mc^2$ with background temperature, which can be justified by the fact that 
the particle is in the presence of a background thermal bath that leads to an increase of its energy when the temperature of the thermal 
bath is increased, according to the factor $\Gamma(T)$\cite{1}. Of course such a temperature would be increased if the radius of the
universe decreases. 

\section{Invariance of the fine structure constant with the cosmological time}

 Now, let us consider the energy of a photon modified by the presence of a given background temperature, namely: 

\begin{equation}
E(T)=p^{\prime}c^{\prime},
\end{equation}
where $p^{\prime}$ represents the modified momentum of the photon. 

On the other hand, we already know that the energy of a photon is $E=h\nu=\hbar w$, where $\hbar=h/2\pi$ and $w=2\pi\nu=2\pi c/\lambda$,
$\lambda$ being the wavelength of the photon and $\nu(=c/\lambda)$ being its frequency. Now, if we consider its energy modified 
by the background temperature, we find 

\begin{equation}
E(T)=\frac{E}{\left(1-\frac{T}{T_P}\right)}=h^{\prime}\frac{c^{\prime}}{\lambda},
\end{equation}

By introducing Eq.(6) ($c^{\prime}$) into Eq.(9), we write

\begin{equation}
E(T)=\frac{E}{\left(1-\frac{T}{T_P}\right)}=h^{\prime}\frac{c}{\lambda\sqrt{1-\frac{T}{T_P}}}=h^{\prime}\frac{\nu}{\sqrt{1-\frac{T}{T_P}}},
\end{equation}
where $\nu=c/\lambda$.

In order to recover the usual equation $E=h\nu$ from Eq.(10) above, it is easy to conclude that $h^{\prime}$ should be 
corrected with temperature in the same way of the speed of light $c^{\prime}$ in Eq.(6). So we find

\begin{equation}
 h^{\prime}=h(T)=\gamma_Th=\frac{h}{\sqrt{1-\frac{T}{T_P}}}, 
\end{equation}
or else $\hbar^{\prime}=\hbar(T)=\gamma_T\hbar$, with $\hbar=h/2\pi$. So, in the early universe, when $T=T_P$, we conclude
that $h^{\prime}$ has also diverged like $c^{\prime}$. 
 
By substituting Eq.(11) into Eq.(10), we can simply verify the usual equation $E=h\nu$. 

It is known that $c^2=1/\mu_0\epsilon_0$, where $\mu_0$ is the magnetic permeability of vacuum and $\epsilon_0$ is the electric
permittivity of vacuum. Thus, based on Eq.(6), by correcting this Maxwell relation with the temperature of the universe, we write 

\begin{equation}
c^{\prime 2}=\frac{1}{\mu_0^{\prime}\epsilon_0^{\prime}}=\frac{c^2}{1-\frac{T}{T_P}}=\frac{1}{\mu_0\epsilon_0\left(1-\frac{T}{T_P}\right)},
\end{equation}
from where we extract 

\begin{equation}
\mu_0^{\prime}=\mu_0(T)=\mu_0\sqrt{1-\frac{T}{T_P}}, 
\end{equation}

and 

\begin{equation}
\epsilon_0^{\prime}=\epsilon_0(T)=\epsilon_0\sqrt{1-\frac{T}{T_P}}, 
\end{equation}
since the electric ($\epsilon$) and magnetic ($\mu$) aspects of radiation are in equal-footing.

Now consider two point-like electrons separated by a certain distance $r$. The electric potential energy $U_e(r)$ between them could
be thought in terms of a certain relativistic energy $\Delta m c^2$, so that we have $U_e(r)=\Delta m c^2$. So we write

\begin{equation}
U_e(r)=\frac{e^2}{r}=\frac{q_e^2}{4\pi\epsilon_0r}=\Delta m c^2,
\end{equation}
where $\Delta m$ is a certain relativistic mass related to an electric energy of interaction ($U_e(r)$) between the two electrons. 

Now, by correcting Eq.(15) with the presence of a thermal background field according to Eq.(6), we write the modified electric potential
energy, namely: 

\begin{equation}
U_e(r,T)=\Delta m c^{\prime 2}=\Delta m c(T)^2=\frac{\Delta m c^2}{\left(1-\frac{T}{T_P}\right)}=\frac{e^{\prime 2}}{r},
\end{equation}
from where we get

\begin{equation}
 e^{\prime 2}=e^2(T)=\frac{e^2}{\left(1-\frac{T}{T_P}\right)}, 
\end{equation}

or else

\begin{equation}
 e^{\prime 2}=\frac{q_e^{\prime 2}}{4\pi\epsilon_0^{\prime}}=\frac{q_e^2}{4\pi\epsilon_0\left(1-\frac{T}{T_P}\right)}
\end{equation}

Inserting Eq.(14) into Eq.(18), we find

\begin{equation}
 \frac{q_e^{\prime 2}}{4\pi\epsilon_0\sqrt{1-\frac{T}{T_P}}}=\frac{q_e^2}{4\pi\epsilon_0\left(1-\frac{T}{T_P}\right)}, 
\end{equation}
which implies

\begin{equation}
 q_e^{\prime 2}=\gamma_Tq_e^2=\frac{q_e^2}{\sqrt{1-\frac{T}{T_P}}},
\end{equation}

or

\begin{equation}
 q_e^{\prime}=q_e(T)=\frac{q_e}{\sqrt[4]{1-\frac{T}{T_P}}}
\end{equation}

\begin{figure}
\includegraphics[scale=0.45]{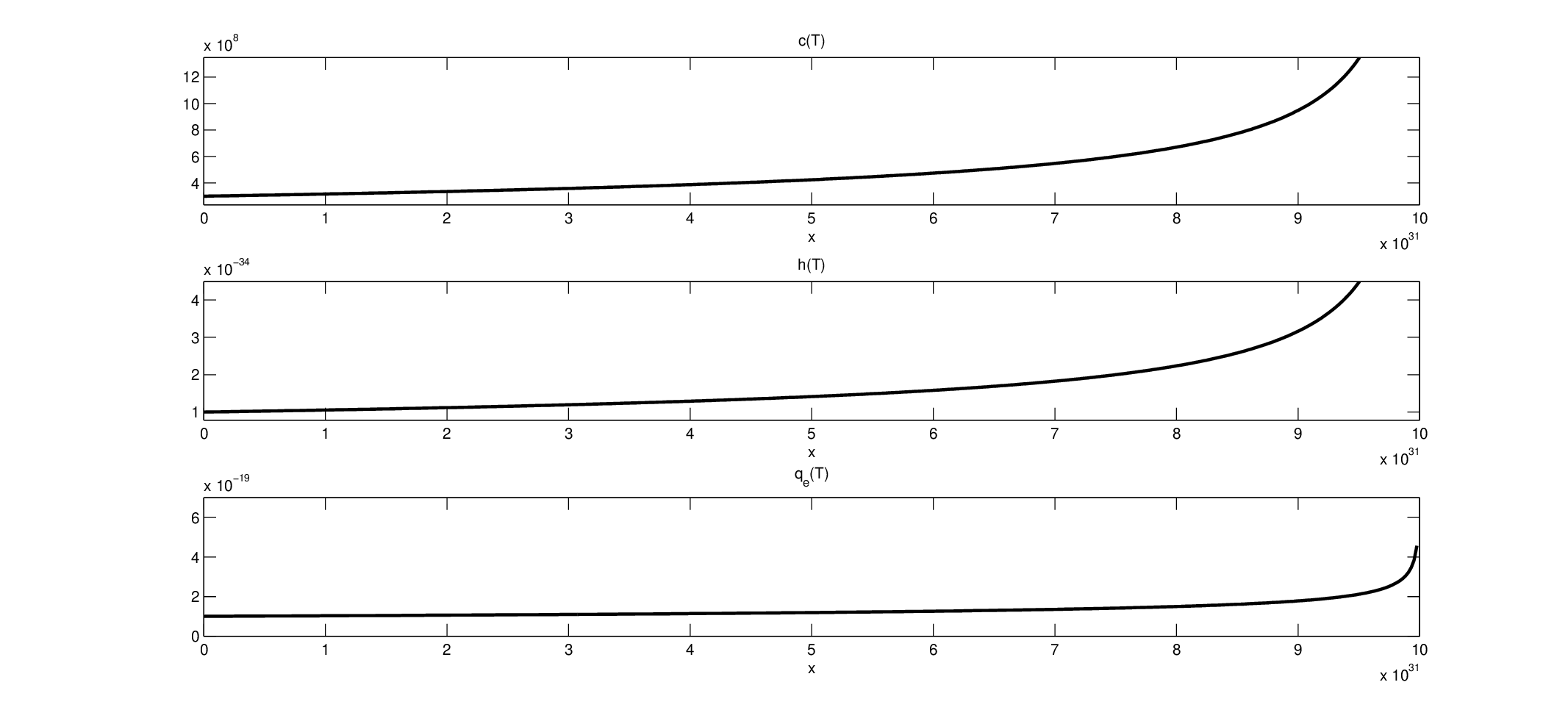}
\caption{The three graphs above provide variations in $c$, $\hbar$ and $q_e$ close to the Planck scale ($\sim 10^{32}$K), 
i.e, only at temperatures of the order of $10^{31}$K, with divergences at $10^{32}$K. Now, it's easy to make a simple extrapolation 
of the graphs and see that, for much lower temperatures, i.e, $T\ll 10^{31}$ K, such constants had too small variations, 
remaining practically constants. Taking the current values ​​of $c\cong 2.9979\times 10^{8}$m/s, 
$\hbar\cong 1.0545\times 10^{-34}$J.s and $q_e\cong 1.6021\times 10^{-19}$C, for example, when computing their values at 
$9\times 10^{31}$K, we find respectively $c^{\prime}\cong 9.4803\times 10^{8}$m/s, $\hbar^{\prime}\cong 2.0953\times 10^{-34}$J.s
and $q_e^{\prime}\cong 2.8491\times 10^{-19}$C. In sum, all these constants varied very rapidly only too close to the Planck temperature,
so that $\alpha$ finally remains invariant\cite{12}.}
\end{figure} 

The fine structure constant without temperature is $\alpha=e^2/\hbar c=q_e^2/4\pi\epsilon_0\hbar c=q_e^2\mu_0 c/2h$. Now,  
by taking into account a given temperature of the expanding universe, we have

\begin{equation}
\alpha(T)=\alpha^{\prime}=\frac{e^{\prime 2}}{\hbar^{\prime}c^{\prime}}=\frac{q_e^{\prime 2}}{4\pi\epsilon_0^{\prime}
\hbar^{\prime}c^{\prime}}=\frac{q_e^{\prime 2}\mu_0^{\prime}c^{\prime}}{2h^{\prime}}
\end{equation}

Finally, by inserting Eq.(6)($c^{\prime}$), Eq.(11)($\hbar^{\prime}$) and Eq.(17)($e^{\prime 2}$) into Eq.(22), or by inserting
$c^{\prime}$, $\hbar^{\prime}$, $\epsilon_0^{\prime}$[Eq.(14)] and $q_e^{\prime 2}$[Eq.(20)] into Eq.(22), or even by inserting
$c^{\prime}$, $h^{\prime}$, $\mu_0^{\prime}$[Eq.(13)] and $q_e^{\prime 2}$[Eq.(20)] into Eq.(22), we find

\begin{equation}
\frac{e^{\prime 2}}{\hbar^{\prime}c^{\prime}}=\frac{e^2}{\hbar c}, 
\end{equation}

or

\begin{equation}
\frac{q_e^{\prime 2}}{4\pi\epsilon_0^{\prime}\hbar^{\prime}c^{\prime}}=\frac{q_e^2}{4\pi\epsilon_0\hbar c} 
\end{equation}

or

\begin{equation}
\frac{q_e^{\prime 2}\mu_0^{\prime}c^{\prime}}{2h^{\prime}}=\frac{q_e^2\mu_0 c}{2h}, 
\end{equation}

that is, 

\begin{equation}
\alpha^{\prime}=\alpha\approx\frac{1}{137.035999037(91)},
\end{equation}
which reveals to us the invariance of the fine structure constant with temperature of the expanding universe and, thus, its invariance
over the cosmic time scale, which is supported by meticulous observational data\cite{12}. 

\section{Conclusions}
 In short, we conclude that, although the universal constants as the speed of light ($c$), the Planck constant ($\hbar$) and the 
electron charge ($e$) have dependence on the temperature of the universe, the fine structure constant $\alpha(=e^{2}/\hbar c\approx 1/137)$
is even more fundamental for remaining invariant with temperature of the expanding universe, probably because $\alpha$ is a dimensionless 
number, since the pure numbers seem to have a special status in the universe.

\end{document}